# Aggregation of P-terphenyl along with PMMA/SA at the Langmuir and Langmuir-Blodgett Films

SYED ARSHAD HUSSAIN[1,2]*, HUGO LEEMAN[2], DEBAJYOTI BHATTACHARJEE[1]
[1]Department of Physics, Tripura University, Suryamaninagar-799130, Tripura India
[2]Centrum voor Oppervlaktechemie en Katalyse, Katholieke Universiteit Leuven, Kasteelpark, Arenberg 23, 3001 Leuven, Belgium.

**Abstract:**

Molecular aggregation and monolayer characteristics of non-amphiphilic p-terphenyl (TP) mixed with either polymethyl methacrylate (PMMA) or stearic acid (SA) at the air-water interface were investigated. The miscibility of the two components was evaluated by measuring and analyzing surface pressure versus area per molecule ($\pi - A$) isotherm. Both attractive and repulsive interactions between the sample (TP) and the matrix (PMMA or SA) were observed depending on the composition and microenvironment. TP and PMMA/SA were not completely miscible in the mixed monolayer. Aggregation and phase separation between sample TP and matrix molecule was revealed by UV-Vis absorption spectroscopic studies and confirmed by scanning electron micrograpgh of LB films.

*email: sa_h153@hotmail.com

## 1. Introduction:

Langmuir-Blodgett (LB) deposition is a potential technique in fabrication of nanofilm of controlled structure [1-3]. Langmuir monolayer at the air-water interface plays an important role in the application of Langmuir-Blodgett (LB) technique [4-6]. The behaviour of Langmuir monolayer at the air-water interface is of great importance for understanding the structure, stability and deposition transfer ratio of the monolayer. By measuring and analyzing the pressure versus area isotherm, one would be able to understand the states of the monolayer and its changes, for example, molecular arrangement, phase structure and phase transition etc. [7-8].

Although amphiphilic molecules are ideally suitable for LB technique, however, recent investigations suggest that non-amphiphilic molecules can also form good quality LB films, when they are mixed with a suitable matrix materials [7-10]. The spectroscopic and aggregating properties of these non-amphiphilic molecules in LB films are very similar to their amphiphilic counter part. It is easy to synthesize the non-amphiphilic molecules and also they are cost effective with respect to their amphiphilic counter part.

Recently mixed monolayers have brought more and more interest in the monolayer research due to the vast properties of the multicomponent system than that of pure ones [11-14]. For some case, monolayer of one component may not be able to transfer onto solid substrate during film deposition and thus second component need to be incorporated into the monolayer to promote the transfer ratio [15-16]. For a mixed monolayer, the miscibility of various components is important to stability and later application of LB films.

In this paper, we report an investigation of the monolayer characteristics and the miscibility of a well known polyphenyl non-amphiphilic p-terphenyl along with stearic acid (SA) or polymethyl methacrylate (PMMA) at the air-water interface. The photophysical characteristics of polyphenyls have been extensively studied and continue to be an interesting area of research owing to the many subtle aspects exhibited by them. The extensive applications of polyphenyls as laser dyes and of their derivatives as excellent liquid crystalline materials [17] make them potentially important for various technological applications. The electronic states of polyphenyls are the functions of the dihedral angles between the benzene rings.

Parameters like temperature, pressure, electric and magnetic field, which affect the dihedral angles between the benzene rings also affect their luminescence spectra. A thermodynamic analysis based on the study of the surface pressure versus molecular area isotherm as a function of the mixture composition allows us to probe the mixing behavior of the two components.

2. **Experimental:**
2.1. **Materials:**
p-terphenyl (98% pure) purchased from Aldrich Chemical Co, USA was purified under vacuum sublimation followed by repeated re-crystallization before use. SA (purity > 99%) purchased from Sigma Chemical Company and isotactic PMMA from Polyscience were used as received. Spectroscopic grade Chloroform (SRL, India) was used as solvent and its purity was checked by fluorescence spectroscopy before use. Solutions of TP, PMMA, SA as well as TP-PMMA and TP-SA mixture at different molefractions were prepared in chloroform solvent and were spread on the water surface.

2.2. **Isotherm measurement:**
A commercially available Langmuir-Blodgett (LB) film deposition instrument (Apex-2000C, India) was used for isotherm measurement. A surface pressure-area per molecule isotherm was obtained by a continuous compression of a monolayer at the air-water interface of the LB trough by a barrier system. The surface pressure at the air-water interface was measured using wilhelmy plate arrangement attached to a microbalance, whose output was interfaced to a microcomputer, which control the barrier movement confining the monolayer at the air-water interface. Milli-Q water was used as subphase and the temperature was maintained at $24^0$C.

Before each isotherm measurement, the trough and barrier were cleaned with ethanol and then rinsed by Milli-Q water. In addition the glass ware was cleaned prior to the use. The surface pressure fluctuation was estimated to be less than 0.5 mN/m during the compression of the entire trough surface area. Then the barrier was moved back to its initial position and the sample containing monolayer forming material was spread on the subphase using a Hamilton microsyringe. After a delay of 30 minutes, to evaporate the solvent, the film at the air water interface was compressed slowly at the rate of 5 mm/min to obtain a single surface pressure versus area per molecule (π-A) isotherm. All isotherms were run several times with freshly prepared solution.

2.3 **UV-Vis absorption spectroscopy:**
UV-Vis absorption spectra were measured using a Parkin Elmer Lambda 25 spectrophotometer.

2.4 **Scanning Electron Microscopy:**
A Hitachi model S-415A electron microscope was used for the SEM imaging of LB films. Accelerating voltage of the electron beam was maintained at 10 kV.

3. **Results and discussions:**
3.1. **Monolayer characteristics of TP at the air-water interface:**
The measurement of the surface pressure versus area per molecule ($\pi - A$) isotherm is the basic technique used for the characterizations of floating monolayer at the air-water interface. The surface pressure ($\pi$) is a measure of the change in surface tension of the subphase covered with a monolayer with respect to pure subphase. The $\pi - A$ isotherm represents the plot of the change of the $\pi$ value as a function of the average area available for one molecule on the subphase surface.

When a dilute solution of pure TP in chloroform solution was spread at the air-water interface and after sufficient time was allowed to evaporate the solvent, the barrier was compressed slowly, surface pressure rises more than 30 mN/m however, large patchy films of TP at the air-water interface was formed which was visible even to the naked eye. Upon relaxation of pressure these large patchy films broke up into several small clusters but did not disintegrate into the molecular level. This suggests that pure TP does not form stable Langmuir monolayer at the air-water interface rather it forms multilayers or clusters of aggregates. Repeated attempt to transfer this floating layer onto solid substrate was failed. However, when a building matrix such as stearic acid (SA) or polymethyl methacrylate (PMMA) was used then the mixed films of TP and SA/PMMA forms excellent and highly stable floating monolayer at the air-water interface and can be easily transferred onto a solid substrate.

Figures 1(a) and 1(b) show the surface pressure (π) versus area per molecule (A) isotherms of TP mixed with PMMA and SA at different molefractions of TP along with pure PMMA, pure SA and pure TP respectively.

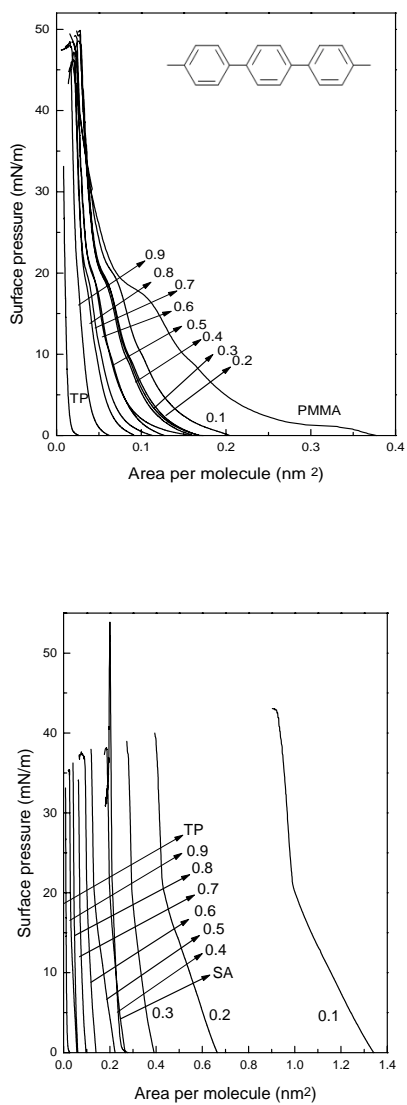

*Fig 1* Surface pressure (π) vs. area per molecule (A) isotherms of TP in (a) PMMA (b) SA matrix at different mole fractions of TP along with pure TP and PMMA/SA isotherm. The numbers denote corresponding mole fractions of TP in PMMA/SA matrix. PM, SA and TP corresponds the pure PMMA, SA and TP isotherm respectively. Inset shows the structure of TP.

The area per molecule of pure PMMA is 0.12 $nm^2$ at a surface pressure of 15 mN/m. In pure PMMA isotherm, there exists certain distinct part with two inflection pints at about 8.2 and 18 mN/m and collapse pressure occurs at about 48.6 mN/m. This is in good agreement with the reported result [18-19]. The transition observed at about 8.2 mN/m is characteristic for isotherm of pure PMMA. The isotherm of PMMA also shows an inflection point at about 20 mN/m, but above this surface pressure, the monolayer remains no longer stable. From the isotherms of TP-PMMA mixed monolayer, it is observed that the inflection portion gradually loses its distinction with increasing molefractions of TP and at 0.7 M and above molefractions of TP, the mixed isotherms show a sharp rise upto a surface pressure of 50 mN/m without any transition point.

It is also interesting to mention in this context that at higher surface pressure, in PMMA matrix, the isotherms of all molefractions of TP almost coincide [figure 1(a)]. This may be due to the fact that at higher surface pressure the mixed monolayer that consists of PMMA matrix remains no longer stable and collapsing of monolayer occurs [19]. Thus at higher surface pressure all the isotherms of TP-PMMA mixed monolayer lose their distinctness and almost overlap on each other.

Pure SA isotherm (figure 1b) is a smoothly increasing curve with lift off area 0.26 $nm^2$ and collapsing of monolayer occurs at about 53.24 mN/m. The area per molecule of pure SA is 0.21 $nm^2$ at a surface pressure of 25 mN/m. These values as well as the shape of the isotherm are well consistent with those values reported in the literature [9-10].

The most interesting feature of TP-SA mixed isotherm is that at lower molefraction of TP upto 0.4 M, the area per molecule of mixed film is larger than that of pure SA. From 0.5 M and above molefraction of TP, the area per molecule is lower than that of SA. The most plausible explanation seems to be that at lower molefraction of TP in SA there may be strong repulsive interaction between TP and SA occurs. However with increasing molefraction of TP, phase separation between TP and SA moeties occurs, which may originate from the immiscibility of the components owing to their differences in molecular structures, physical and chemical properties. Unlike that of the TP-PMMA mixed isotherms TP-SA mixed isotherms do not coincide at higher surface pressure and remained stable until collapsing of monolayer occurs. However, for both the matrices the area per molecule of the mixed monolayer gradually decreases with increase in molefraction of TP. This is an indicative that the TP molecules are successfully incorporated into the matrix materials to form mixed monolayer.

Table 1 shows the monolayer characteristics taken from the isotherms. Mean molecular area or limiting area ($A_0$) occupied per repeating unit in the monolayer. This is estimated by extrapolating the steep rising part of the $\pi - A$ isotherm to zero surface pressure [20].

***Table 1:*** *Monolayer characteristics taken from the $\pi - A$ isotherms*

| Molefraction of TP in PMMA/SA | $A_0$ in PMMA (nm$^2$) | $A_0$ in SA (nm$^2$) | C in PMMA (mN$^{-1}$) | C in SA (mN$^{-1}$) |
|---|---|---|---|---|
| 0 | 0.127 | 0.216 | 36.41 | 4.35 |
| 0.1 | 0.087 | 1.061 | 29.81 | 8.17 |
| 0.2 | 0.07 | 0.452 | 27.1 | 12.05 |
| 0.3 | 0.069 | 0.313 | 26.7 | 7 |
| 0.4 | 0.07 | 0.211 | 26.78 | 7.27 |
| 0.5 | 0.05 | 0.145 | 26.01 | 13.9 |
| 0.6 | 0.048 | 0.107 | 26.1 | 8.73 |
| 0.7 | 0.045 | 0.076 | 24.97 | 8.13 |
| 0.8 | 0.042 | 0.049 | 21.67 | 7.67 |
| 0.9 | 0.033 | 0.034 | 18.27 | 16.4 |
| 1 | 0.013 | 0.013 | 14.11 | 14.11 |

### 3.2. Apparent Compressibility:

To obtain more information from the $\pi - A$ isotherms, the apparent compressibilities of the films are calculated by using equation 1 and listed in table 1.

The apparent compressibility C is defined as [21-22]

$$C = -\frac{1}{A_{10}} \frac{A_{30} - A_{10}}{30 - 10} \quad (1)$$

Where $A_{10}$ and $A_{30}$ are the areas per molecule at 10 and 30 mN/m surface pressure respectively.

The compressibility of pure PMMA film is 36.41 mN$^{-1}$. This is consistent with the distinct segment observed in PMMA isotherm (figure 1a). Although with the increase in molefraction of TP in PMMA matrix the C value decreases but interestingly the compressibility of the mixed monolayer of 0.3 – 0.6 M of TP in PMMA are almost same. Interestingly in SA matrix for 0.3, 0.4, 0.6 – 0.8 M of TP the C value is also almost constant although the overall C value for TP-SA mixed films are lower. This may be the characteristics of the sample.

The low values of compressibility makes the TP-SA mixed films more harder with respect to TP-PMMA mixed films. This is reasonable that fatty acid monolayer on water surface has the closest molecular packing and can be considered as a two dimensional crystal solid [21].

### 3.3. Miscibility of TP with the matrix molecules:

The compression isotherm of a two dimensional Langmuir film as a plot of surface pressure ($\pi$) versus area per molecule (A) whose surface pressure is given by $\pi = \gamma_0 + \gamma$, where $\gamma_0$ and $\gamma$ are the surface tension of the clean subphase and the subphase covered with film material respectively.

An ideally mixed monolayer and a completely immiscible monolayer are absolutely opposite to each other. In an ideal mixed monolayer, the intermolecular attractive forces (F) between the constituent molecules 1 and 2 is given by $F_{11} = F_{12} = F_{22}$, whereas for a completely immiscible monolayer $F_{11} \rangle F_{12} \langle F_{22}$. Where $F_{ij}$ are the attractive forces between molecules of the two components $i$ and $j$ ($i, j$ = 1, 2 respectively).

In terms of the molecular area of the mixed monolayer, the ideal value of the molecular area ($A_{id}$) is calculated from the molar ratio of the two components as

$$A_{id} = A_1 N_1 + A_2 N_2$$

Where $N_1$ and $N_2$ are the molar fraction of the pure components in the mixture, $A_1$ and $A_2$ are the area occupied by the monomer of the pure components of the constituent molecules.

The excess area, $A_E$, which is the difference in molecular area between the ideal ($A_{id}$) and the experimentally observed molecular area ($A_{12}$) and is given by the following relation

$$A_E = A_{12} - A_{id}$$

$A_E$ is the measure of the interaction between the constituent components in the mixed monolayer, when there is an interaction between the constituent molecules in the mixed monolayer the experimentally observed molecular area ($A_{12}$) deviates from the value supposed for ideally mixed monolayer. Once the two components form an ideally mixed monolayer or they are immiscible, the $A_{id}$ will equalize $A_{12}$. That is if an ideally mixed monolayer is formed and the constituent components are completely immiscible, the excess area ($A_E$) will be zero, i.e., $A_E = A_{12} - A_{id} = 0$. The plot of $A_{12}$ versus $N_1$ will give a straight line. Any deviation from the straight line indicates miscibility and non-ideality [21-23]. Actually there are always intermolecular forces between the constituent molecules in the mixed monolayer. The magnitude of these forces determine the $A_E$. For a mixed monolayer, $A_E$ will be negative if the intermolecular forces are attractive i.e., greater cohesive forces between the unlike components exists on the other hand $A_E$ will be positive for the repulsive interaction between the constituent molecules i.e., greater cohesive forces between the like components.

Figure 2a and 2b show the plot of the data of the actual area ($A_{12}$) per molecule versus mole fraction of TP in the mixed monolayer with PMMA and SA respectively at different fixed surface pressure of 5, 10, 15, 20, 25 and 30 mN/m.

The mixing behaviour of TP in PMMA matrix (figure 2a) is quite interesting. At surface pressure of 5, 10 and 15 mN/m there is a negative deviation from the ideality curve for lower molefraction (0.1 – 0.5 M of TP) is observed, whereas, for higher mole fraction the experimentally observed values almost coincide with the ideality. This is an indication of certain miscibility or nonideality and the cohesive force between TP and PMMA predominates at lower mole fractions of TP in PMMA. Whereas, with the increase in mole fraction the TP and PMMA are completely immiscible resulting an ideal mixed monolayer at the air-water interface.

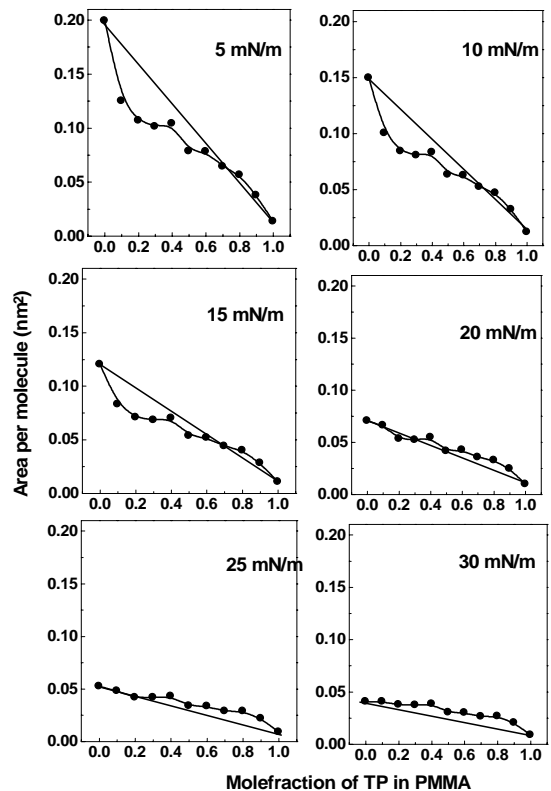

**Fig 2 (a)** The plot of the data of the actual area ($A_{12}$) per molecule versus mole fraction of TP in the TP-PMMA mixed monolayer for different fixed surface pressure. The numbers denote the surface pressures.

At higher surface pressures of 20, 25 and 30 mN/m ideal mixing of TP and PMMA occurs for lower mole fraction of TP and at higher mole fraction slight positive deviation from the ideal behaviour is observed. The repulsive interaction between TP and PMMA and cohesive force between TP-TP and PMMA-PMMA may be responsible for this positive deviation at higher mole fraction of TP at higher surface pressures.

In SA matrix the experimental data show positive deviations from the ideality curve for lower molefraction up to 0.5 M. The positive excess areas in case of SA matrix as discussed previously is an indication of more favourable cohesive interactions between the like molecules (TP-TP and SA-SA) than the unlike components (TP-SA) which can lead to partial or total phase separation and formation of clusters or microcrystalline aggregates of TP molecules in the SA-mixed LB films. However for higher molefraction of TP the sample and matrix molecules are immiscible and ideally mixed monolayer is formed.

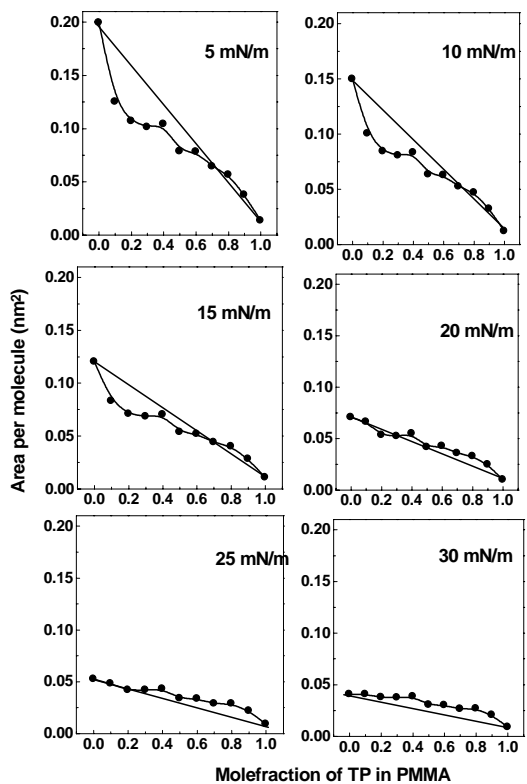

*Fig. 2 (b) The plot of the data of the actual area ($A_{12}$) per molecule versus mole fraction of TP in the TP-SA mixed monolayer for different fixed surface pressure. The numbers denote the surface pressures.*

Figure 3 shows the plot of the collapse pressure versus mole fraction of TP in PMMA and SA matrix. According to the phase rule [21], the two components in the mixed monolayer's are miscible at the interfaces when the collapse pressure varies with the compositions of the mixed monolayer. In the present case the collapse pressures are almost independent of the mole fractions except a small variation. The appearance of the phase transition points similar to that of pure components in the mixed monolayer also suggest the existence of pure SA/PMMA phase in the mixed monolayer. That is the two components TP and PMMA/SA are not completely miscible in the mixed monolayer.

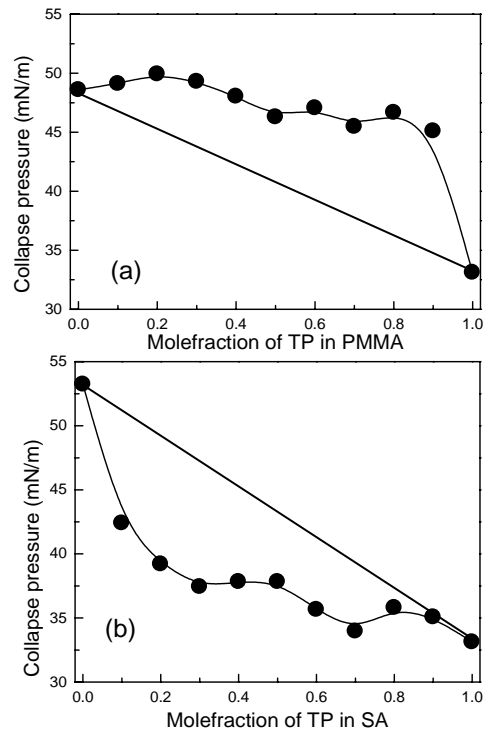

*Fig. 3 Collapse pressures versus molefraction plot for (a) TP-PMMA and (b) TP-SA mixed films.*

### 3.4 UV-Vis absorption spectroscopy:

Figure 4 shows the UV-Vis absorption spectra of LB films of TP along with the solution and microcrystal spectrum.

TP solution absorption spectrum possess a distinct broad and structureless band in the 200–325 nm region with peak at around 280 nm. This is consistent with the literature reported elsewhere [24]. This band has been assigned to $^{1}L_{a}$ state [24]. Also there is a almost indistinguishable weak hump is observed at about 240 nm in the solution absorption spectrum. The microcrystal absorption spectrum is quite different with respect to TP solution absorption spectrum. Unlike that of the solution spectrum microcrystal spectrum possess distinct band system within 200-300 nm region with specific structural features. The high-energy band is located at 229 nm along with a broad low intense band in the longer wavelength region with diffused but prominent vibrational bands with peaks at 261, 275 and 285 nm, respectively, which are almost indistinguishable in absorption spectra of LB films. The absorption spectra of mixed LB films for both matrices show almost similar band profile and have distinct similarity with the microcrystal absorption spectrum with high-energy band at around 233 nm and a broad low intense band in the longer wavelength region with peak at around 290 nm. In LB film spectra the vibrational peaks are not so prominent. The weak hump at 240 nm in solution absorption spectrum give rise to the intense prominent high-energy band at around 233 nm in the LB films. Moreover, all the absorption bands in the LB films are overall broadened. This may be due to the change of planarity of TP molecules due to the organization in the mixed LB films when goes from solution to films/solid states.

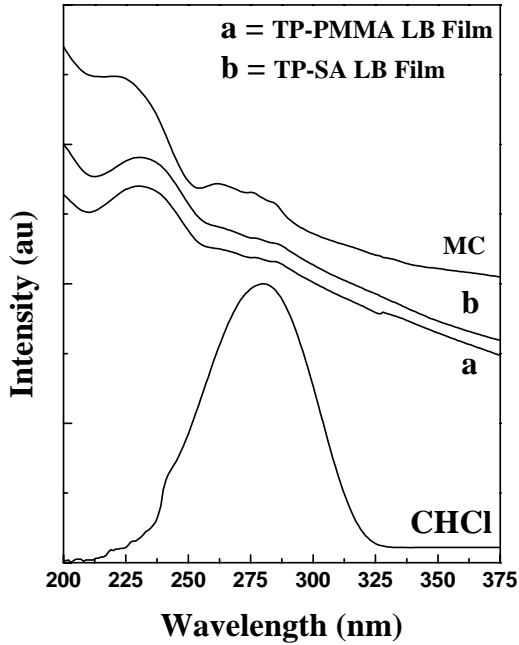

*Fig. 4* UV-Vis absorption spectra of TP solution, microcrystal and LB films of 0.5M TP in PMMA and SA.

In this context it is interesting to mention that molecular exciton coupling theory [25] predicts that in the simplest case of a two molecule system, the dipole-dipole interactions result in splitting the energy level of the excited state into two levels with higher and lower energies relative to the undisturbed excited state. The change in energies of these two levels is Davydov splitting and can be calculated from the equation:

$$\Delta E = E'' - E' = 2 \cdot \left( \frac{M_i . M_j}{R_{ij}^3} - 3 \frac{(M_i . R_i).(M_j . R_j)}{R_{ij}^5} \right)$$

Here, $M_i$ and $M_j$ are the electric dipole transition moments in molecules $i$ and $j$, respectively, and $R_{ij}$ is the centre to centre distance between these molecules. In the case of the co-planar alignment of the dipole transition moments of two molecules, the transition to one of the excited states, corresponding to the antiparallel arrangement of dipole moments is forbidden. The energy difference between the excited monomer level and the exciton level, $\Delta \varepsilon$, is given by [25]:

$$\Delta \varepsilon = \frac{1}{2} \Delta E = \frac{M^2}{R_{ij}^3}(1 - 3\cos^2 \theta)$$

Here, $\theta$ is the angle between the transition moment direction and the line connecting the molecules' centres. When $0^0 \leq \theta < 54.7^0$, the exciton level is energetically located below the monomer level causing a red shift in the electronic absorption spectrum, creating aggregates termed J-aggregates [26]. For $54.7^0 < \theta \leq 90^0$, the exciton level is located energetically above the monomer level, causing a blue shift. Corresponding aggregates are called H-aggregates [26]. When $\theta = 54.7^0$, no shift in the absorption spectrum is observed. The aggregates are then termed I-aggregates [27].

So in the present case the observed blue shift accompanied with the broadening of the band profile in LB films with respect to the solution spectrum may be due to the aggregation of TP molecules in the films. In the light of the molecular excitation theory this aggregation is identified as H-aggregates. Our SEM studies also support the aggregation of the TP molecules due to the phase separation between the samples and the matrix molecules.

**3.5 Scanning electron micrograph:**
In Figure 5 we have shown the SEM picture of the 10 bilayer LB film (TP:SA = 1:1). Under an optical microscope the LB film appeared homogeneous but the SEM picture reveals a clear heterogeneous morphology. The SEM image of TP-PMMA mixed LB films also shows similar morphology (figure not shown) The homogeneous dark background is due to of SA forming larger region compared to TP, and TP molecules seems to be in islands above that. Strong repulsive interaction between TP and matrix molecule may squeeze out TP molecules in matrix. This is expected since the nature of TP and the matrix molecules are quite different, which is reflected in the SEM morphology. The formation of aggregates due to the phase separation and demixing of binary components in LB film is thus confirmed.

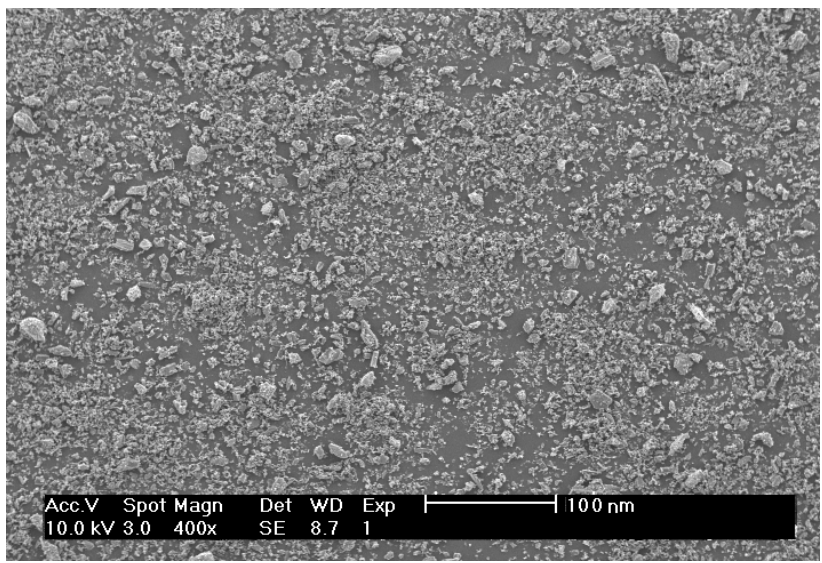

*Fig. 5* *Scanning electron micrograph of mixed LB film (SA:TP = 1:1, 10 bilayer)*

### 4. Conclusion:

Although non-amphiphilic TP molecules are not ideally suited for LB technique, however, our investigation showed that stable self supporting Langmuir monolayer of TP can be formed when they are mixed with either PMMA or SA. TP-PMMA mixed films may not be stable at higher pressure but the TP-SA mixed films were stable until collapse pressure was reached. The compressibility of TP-SA mixed films were observed to be lower than the TP-PMMA mixed films. The interaction between sample (TP) and matrix (PMMA/SA) molecules largely depends on the mole fraction of mixing and the surface pressures. At lower surface pressure and lower mole fraction there exist an attractive interaction between TP and PMMA, whereas, repulsive interaction predominates at higher surface pressure and higher mole fractions. In SA matrix strong repulsive interaction between TP and SA and strong cohesive interaction between the TP-TP and SA-SA were observed at all surface pressures for lower mole fraction of TP. Whereas, at higher mole fraction of TP the sample and matrix molecules are immiscible and form ideal mixed monolayer. H-type of aggregation of TP molecules in the mixed LB films was revealed by UV-Vis absorption spectroscopy. SEM studies confirmed the phase separation and demixing between sample TP and the matrix molecules.